\documentstyle[12pt]{article}
\def\be{\begin{equation}}
\def\ee{\end{equation}}
\def\ba{\begin{array}}
\def\ea{\end{array}}
\def\ni{\noindent}
\newcommand{\scs}[1]{{\scriptscriptstyle #1}}
\newcommand{\eqr}[1]{(\ref{#1})}
\def\L{\Lambda}
\def\a{\alpha}
\def\pint{-\!\!\!\!\!\!\int}
\def\pinta{-\!\!\!\!\!\!\int_a^b}
\def\pintak{-\!\!\!\!\!\!\int_{b_k}^{a_{k+1}}}

\newcommand{\N}{\scs N}
\begin{document}
\begin{titlepage}
\begin{flushright}CERN-TH.7390/94\\TAUP-2191-94\end{flushright}
\vskip1.5cm
\begin{center}
{\huge\bf Large N phase transitions}\\
{\huge\bf and multi-critical behaviour}\\
\vskip.2cm
{\huge\bf in generalized 2D QCD}\\
\vskip2cm
{\Large B. Rusakov$^*$ and S. Yankielowicz$^{**}$
\footnote{Supported in part by the US--Israel Binational Science
Foundation, `GIF' -- the German--Israeli Foundation for Scientific
Research and Development and the Israel Academy of Science.} }\\
\vskip.3cm
          {\em $^*$ICTP, High Energy Section,\\
          P.O.Box 586, I-34014 Trieste, Italy}\\
\vskip.3cm
          {\em $^{**}$School of Physics and Astronomy \\
          Raymond and Beverly Sackler Faculty of Exact Sciences \\
          Tel-Aviv University, Tel-Aviv 69978, Israel \\
          and \\
          Theory Division, CERN \\
          CH-1211 Geneva 23, Switzerland \\}

\vskip1.5cm

{\large\bf Abstract.}
\end{center}
Using matrix model techniques we investigate the large N limit
of generalized $2D$ Yang-Mills theory. The model has a very rich
phase structure. It exhibits multi-critical behavior and reveals
a third order phase transitions at all genera besides {\it torus}.
This is to be contrasted with ordinary $2D$ Yang-Mills which,
at large N, exhibits phase transition only for spherical topology.

\vskip1.5cm

\ni
CERN-TH.7390/94\\
August 1994

\end{titlepage}\vfill\eject

\newpage

Recently, generalized $2D$ pure gauge SU(N) and U(N) models (gYM$_2$)
have been proposed \cite{DKS} and solved at arbitrary N \cite{TAU}.
The solution generalizes the one obtained for ordinary $2D$ Yang-Mills
model (YM$_2$) \cite{Mig}-\cite{Witt}.

In particular, the partition function for a closed surface of genus
$g$ and  area $A$ (coupling constant absorbed into area) has the form:
\be Z_g(A)=\sum_r d_r^{2(1-g)}\exp\Big(-\frac{A}{2N}\L(r)\Big)
\label{gZ}\ee
Here, $r$ is an irreducible representation of the gauge group,
which is parametrized by its highest weight components,
$n_1\ge ...\ge n_\N$, associated with the lengths of lines in the
Young table; $d_r$ is its dimension,
\be d_r = \prod_{1\leq i < j \leq N} \Big(1+\frac{n_i-n_j}{j-i}\Big)
\label{dim}\ee and \be\L(r)=N\sum_{k>0}\frac{a_k}{N^k}C_k(r)
\label{new}\ee with $C_k(r)$ being the $k$-th Casimir operator
eigenvalue, \be C_k(r)=\sum_{i=1}^N l_i^k\gamma_i \;\;,\ee
where \be\gamma_i=\prod_{j\neq i}\Big(1-\frac{1}{l_i-l_j}\Big)
\hskip1.5cm  l_i=n_i+N-i\;.\ee

Formula \eqr{gZ} generalizes the known result \cite{Rus1},\cite{Witt}
for YM$_2$ which corresponds to $\L(r)$ with only the $k=2$ term
present. The eigenvalue of the second Casimir operator is
\be C_2(r)=\sum_{i=1}^{N}n_i(n_i+N+1-2i)\;\;.\label{C2}\ee
A generalization of the result for Wilson loop averages \cite{Rus1}
is,
\be W_g(C)=\sum_{r_1,...,r_m}\Phi_{r_1...r_m}\prod_{k=1}^m
d_{r_k}^{2(1-g_k)}\exp\Big(-\frac{A_k}{2N}\L(r_k)\Big)\;\;,
\label{loop}\ee
where $m$ is the number of windows, $A_k$'s are the corresponding areas,
$g_k$ is the ``genus per window" and the coefficients $\Phi_{r_1...r_m}$
are the U(N) (SU(N)) group factors which depend on the contour topology
(see Ref.\cite{Rus1} for details). The result depicted in \eqr{loop}
is obtained from the YM$_2$ result
by the replacement $C_2(r)\to\L(r)$

In Ref. \cite{TAU} the large N expansion of the generalized action was
carried out and a stringy description was derived. The partition
function \eqr{gZ} can be written as a sum of $2D$ string maps, where
maps with branch points of degree higher than one as well as
``microscopic surfaces" play an important role. This generalizes the
string interpretation of YM$_2$ given previously by Gross and Taylor
\cite{Gross}-\cite{GT2}. It is still a challenging problem to find the
string action which gives rise to the sum of maps that reproduces the
gauge theory partition function. Recently, an important progress toward
this goal has been made \cite{Hor},\cite{CMR}. The results of
Ref.~\cite{CMR} reveal the topological aspects of the underlying string
theory associated with holomorphic maps.

\bigskip

In the present letter we continue to study the gYM$_2$ theory.
We employ the large N approach, elaborated in details in \cite{Rus2},
which is well known within the context of matrix models. We demonstrate
that gYM$_2$ exhibits a much richer structure than YM$_2$.
In particular,
besides the phase transition realized \cite{DK} in YM$_2$ on a sphere,
there are phase transitions at any genus in gYM$_2$. Moreover,
at any genus the model accommodates multi-critical behaviour.

Following \cite{Rus2} we write the partition function \eqr{gZ} at
large N as a path integral over continuous Young tables. Namely, we
introduce the continuous function,
\be h(x)=\lim_{N\to\infty}\frac{1}{N}\Big(k-\frac{N}{2}-n_k\Big)\;\;,
\hskip1cm x=\frac{k}{N}\;.\ee
Then, the partition function takes the form:
\be Z=\int\prod_{0\leq x\leq 1} dh(x) e^{S[h(x)]}\;\;,\label{Z}\ee
\be S=\frac{N^2}{2} \int_0^1\!\!\! dx\left\{
-AV[h(x)]+(2-2g)\;-\!\!\!\!\!\!\int_0^1 dy\log\Big|h(x)-h(y)\Big|
\right\},\label{act}\ee where $$V(h)=\sum_{k>0}\a_k h^k\;\;.$$
It is implied that the path integral \eqr{Z} should be taken over
$h(x)$'s which satisfy the constraint $$ \frac{dh}{dx}\geq 1 $$
which is reminiscence of the dominance condition for group
representations.

Then, the saddle point equation is
\be\frac{1}{2}\xi V'_h[h(x)]=\pint\frac{dy}{h(x)-h(y)}\;,\hskip1cm
\xi=\frac{A}{2-2g}\label{sp}\ee
($\xi^{-1}$ is the surface density of Euler characteristics).

Introducing the density $$\rho(h)=\frac{dx}{dh}$$
which should be positive and normalized to
\be\int_{a}^{b}\!\! dh\;\rho(h)=1\;,\label{norm}\ee
we rewrite \eqr{sp} as
\be \frac{1}{2}\xi V'(h)=\pinta\frac{dx\rho(x)}{h-x}\;,
\hskip1cm a\leq h\leq b \; . \label{SP}\ee

The constraint $h'(x)\geq 1$ now takes the form
\be\rho(h)\leq 1\;\;.\label{constr}\ee

The solution of \eqr{SP} is
\be\rho(h)=\frac{1}{\pi}P(h)\sqrt{(h-a)(b-h)}\label{rho_weak}\;\;,\ee
where the polynomial $P(h)$ and the values $a$ and $b$ are uniquely
defined by the $h\to\infty$ asymptotic behavior
\be\frac{1}{2}\xi V'(h)-P(h)\sqrt{(h-a)(h-b)}\sim
\frac{1}{h}\;\;.\ee

In the YM$_2$ case ($V=h^2$), the solution (constraint \eqr{constr}
is ignored) is:
\be\rho(h)=\frac{\xi}{\pi}\sqrt{b^2-h^2}\hskip1cm |h|\leq 1
\hskip1cm \xi b^2=2\label{old_weak}\;\;.\ee

Thus, the solution exists only for genus zero, i.e., the sphere.
At $2\xi<\pi^2$ Eq.\eqr{constr} is satisfied for all $h$.
It means that in sum over representations we can ignore the dominance
condition and all $n_k$'s now run independently from $-\infty$
to $+\infty$. In other words the U(N) (as well as SU(N)) model now
becomes (effectively) $\bigotimes^N$U(1) model, i.e., abelian.
In terms of the original U(N) matrix variables it means that all
unitary matrices are frozen near $I$.
This is the weak coupling (small area) phase of the model.

For $2\xi>\pi^2$, there is a region of values of $h$ where solution
\eqr{old_weak} does not satisfy the inequality \eqr{constr}.
In this region the inequality \eqr{constr} is saturated, $\rho(h)=1$.
Then, the solution of the saddle point equation is \cite{DK}:
\be\rho(h)=\left\{\ba{ll} -\frac{2}{\pi ah}\sqrt{(a^2-h^2)(h^2-b^2)}
\Pi_1\Big(-\frac{b^2}{h^2},\frac{b}{a}\Big) & \mbox{ for $-a<h<-b$}\\
1 & \mbox{ for $-b<h<b$}\\
\frac{2}{\pi ah}\sqrt{(a^2-h^2)(h^2-b^2)}
\Pi_1\Big(-\frac{b^2}{h^2},\frac{b}{a}\Big) & \mbox{ for $b<h<a$}
\ea\right. \label{old_strong}\ee
where $\Pi_1(x,k)$ is the complete elliptic integral of the third kind
with modulus $k=\frac{b}{a}$ and the parameters are determined by the
equations \be a(2E-k'^2K)=1 \hskip1.5cm aA=4K\;\;,\ee
where\footnote{All notations are taken from the book \cite{Erd}.}
$k'=\sqrt{1-k^2}$ is the complementary modulus and
$$K=K(k)=\int_0^1\frac{dx}{\sqrt{(1-x^2)(1-k^2x^2)}}\hskip1cm
E=E(k)=\int_0^1 dx\sqrt{\frac{1-k^2x^2}{1-x^2}}\;.$$
At the critical value, $2\xi_c=\pi^2$ ($A_c=\pi^2$), the third order
phase transition \cite{GW} takes place. It was observed by Douglas
and Kazakov \cite{DK}.

We observe that in the generalized case the constraint \eqr{constr}
can be satisfied also for negative $\xi$ by an appropriate choice of
the coupling constants $\a_i$, which implies the existence of
phase transitions at higher genera, except the {\it torus}. (In the
latter case the saddle point equation is an algebraic equation for $h$
with solution
$h=const$ and, therefore, the constraint $h'\geq 1$ is not satisfied).
Moreover, there is a multi-critical behaviour at each genus. Indeed,
since the semicircle distribution \eqr{old_weak} is now deformed by a
polynomial \eqr{rho_weak}, there are several disconnected regions of $h$
where the constraint \eqr{constr} is satisfied.

Consider, for example, the simplest generalized case
$$V(h)=h^2+\frac{2\a}{3} h^3$$ which is the well-known $\varphi^3$
model. The solution is \cite{BIPZ}:
\be \rho(h)=\frac{\xi}{\pi}[1+\frac{\a}{2}(a+b)+\a h]
\sqrt{(h-a)(b-h)}\label{cub}\;\;,\ee
and the values of $a$ and $b$ are determined from
\be\ba{lll} &\frac{\a}{4}(b-a)^2+(a+b)[1+\frac{\a}{2}(a+b)]=0\\
&\\         &\xi(b-a)^2[1+\a(a+b)]=8\;\;, \ea\ee
which leads to
\be\ba{lll} &a=\frac{2z-1}{2\a}-2\sqrt{\frac{z}{\xi}}\\
&\\         &b=\frac{2z-1}{2\a}+2\sqrt{\frac{z}{\xi}}\;\;, \ea\ee
with \be z^3-\frac{1}{4}z+\frac{\a^2}{2\xi}=0\;\;.\ee

At negative $\xi$ we should take the $z<0$ solution which exists for
$$\a^2<-\frac{\xi}{3\sqrt{12}}\;\;.$$

The critical value $\xi_c=\xi_c(\a)$ should be defined from \eqr{cub}
through the condition \be\rho_{\rm max}(h)=1.\label{crit}\ee

It is clear that similar situation prevails for the general case.
For $\xi<\xi_c$, where $\xi_c$ is defined by \eqr{crit} and
$\rho$ defined by \eqr{rho_weak}, the weak coupling phase is realized.

In the strong coupling
phase ($\xi>\xi_c$) the saddle point equation \eqr{SP}
should be solved in presence of the constraint \eqr{constr}. The only
solution is
\be\rho(h)=\left\{\ba{lll}&1\hskip1.5cm a_k\leq h\leq b_k\hskip1.5cm
k=1,...,m\\&\\&
\tilde{\rho}(h)\hskip1cm b_k\leq h\leq a_{k+1}\hskip1cm k=0,...,m
\ea\right.\label{rho_strong}\ee
where $b_0=a$, ~ $a_{m+1}=b$ and $m$ is the number of intervals
($a_k,b_k$) where
$$\frac{1}{\pi}P(h)\sqrt{(h-a)(b-h)}>1\;\;.$$

The function $\tilde{\rho}(h)$ has to be defined via the equation
\be \frac{1}{2}\xi V'(h)-\sum_{k=1}^m\log\frac{h-b_k}{h-a_k}=
\sum_{k=0}^m\pintak\frac{dx\tilde{\rho}(x)}{h-x}\;\;.
\label{rho_tild}\ee

\bigskip

Wilson loop averages can be calculated following the same steps as
in the YM$_2$ case on a sphere \cite{Dima},\cite{Rus3}. A simple
(without self-intersections) contractible
\footnote{An expectation value of non-contractible loop
might be zero due to pure topological reason. A complete set
of zero-valued loops was found in \cite{Rus1}.}
loop divides the surface into two ``windows" with areas $A_1,A_2$
and genera per ``window" $g_1,g_2$. The corresponding expectation
value can be calculated in a similar way to
what was done for YM$_2$ on a sphere \cite{Dima}:
\be W(A_1,g_1;A_2,g_2)= \int dh\eta(h)D^{2-2g_2}(h)
e^{-\frac{1}{2}A_2V'(h)}\;\;,\label{W}\ee
where the quantities $\eta(h)$ and $D(h)$ are precisely the same
as in the spherical case (see Ref.\cite{Dima}) and are expressed
through $\rho(h)$. We note that Eq.\eqr{W} is symmetric
with respect to the exchange $(A_1,g_1)\to (A_2,g_2)$. Again, as in
the spherical case, $W(A_1,0)=W(0,A_2)=1$.

Expression \eqr{W} is valid in both the weak and the strong coupling
phases. A complete information concerning the phase transition is
contained in the function $\rho(h)$. This function is given by
\eqr{rho_weak} in the weak coupling phase and by
\eqr{rho_strong}-\eqr{rho_tild} in the strong coupling phase.

Starting from \eqr{W} the complete set of Wilson loop averages
can be calculated by means of the loop equation \cite{MM} in its
simple two-dimensional formulation \cite{KK},
similarly to what was done for YM$_2$ on a sphere \cite{Rus3}.

\bigskip

Several remarks are in order. First, the non-trivial saddle point
for $g>1$ exists only if the coupling constant corresponding to
the highest Casimir eigenvalue in the generalized potential \eqr{new}
is negative. In terms of the matrix model it implies that the potential
$V(h)$ has a negative leading term, i.e., we deal with a matrix model
with an {\em upside-down} potential. It is easy to check that there is
no phase transition on a sphere for the case of an upside-down potential.
Our second remark concerns the function $\xi_c=\xi_c(g)$. For any
particular model (fixed couplings $\a_k$ and maximal order of the
Casimir operators entering in \eqr{new}) the critical value $\xi_c$ is
fixed (it is unambiguously defined by the saddle point equation).
Therefore, increasing the genus the critical area increases
proportionally in order to keep fixed the critical number of handles per
area (the density of the Euler characteristics). Also, if we fix the
area $A$ of the surface, then with the increase of the genus the number
of multi-critical points decreases
and for $g>1+\frac{A}{2|\xi_c|}$ there is no phase transition.

There are several open questions which follow from our observations.
We have presented only a qualitative picture
of the critical behavior in gYM$_2$. It is of interest to get
the critical exponents associated with the phase transitions.

In the weak coupling phase, where the constraint \eqr{constr} can be
ignored, the model coincides with the usual (large N) hermitian matrix
model with arbitrary potential. The latter has been investigated
thoroughly during recent years. It describes (upon fine tuning of the
parameters) two-dimensional quantum gravity (coupled to non-unitary CFT
matter). It can be also viewed as string embedded in $D=0$ space. It is
very intriguing to find out whether there exists an hermitian matrix
model which corresponds to the strong coupling phase of gYM$_2$. If such
a matrix model description is found, then the obvious question is to
which quantum gravity string theory it corresponds and what is its
relation to the stringy formulation of gYM$_2$
\cite{TAU},\cite{Gross}-\cite{CMR}.
Clearly, the stringy approach to gYM$_2$ holds in the strong coupling
phase. As a first step toward the construction of such a matrix model
one could try to introduce the constraint \eqr{constr} directly into
the action \eqr{act} of the model (the constraint can be included in
the path integral over $h(x)$ as a step function $\theta(|h'(x)|-1)$
which can be incorporated into the action, for example,
via a Lagrange multiplier or through some convenient parametrization).

One can start with the lattice formulation of gYM$_2$ and integrate
systematically over the link gauge variables. This gives rise to an
effective action for the auxilliary B field which is again a matrix
model. The matrix model formulation can help toward finding an
integrable structure of gYM$_2$. \footnote{We thank O. Ganor for
collaborating on this point.}

\bigskip

\ni
{\large\bf Acknowledgments.}\\
\ni B.R. would like to thank High Energy Group of Tel-Aviv University
for kind hospitality.

\end{document}